# Strong bulk spin-orbit torques quantified in the van der Waals ferromagnet Fe$_3$GeTe$_2$

Franziska Martin[1], Kyujoon Lee[1,2,‡], Maurice Schmitt[1], Anna Liedtke[1], Aga Shahee[1], Haakon Thømt Simensen[3], Tanja Scholz[4], Tom G. Saunderson[1], Dongwook Go[5], Martin Gradhand[1,6], Yuriy Mokrousov[1,5], Thibaud Denneulin[7], András Kovács[7], Bettina Lotsch[4], Arne Brataas[3], Mathias Kläui[1,3]

[1]Institute of Physics, Johannes Gutenberg-University, 55099 Mainz, Germany
[2]Division of Display and Semiconductor Physics, Korea University, 30019 Sejong, Korea
[3]Centre for Quantum Spintronics, Department of Physics, Norwegian University of Science and Technology, 7491 Trondheim, Norway
[4]Max Planck Institute for solid state research, 70569 Stuttgart, Germany
[5]Peter Grünberg Institut and Institute for Advanced Simulation, Forschungszentrum Jülich and JARA, 52425 Jülich, Germany
[6]H. H. Wills Physics Laboratory, University of Bristol, Bristol BS8 1TL, United Kingdom
[7]Ernst Ruska-Centre for Microscopy and Spectroscopy with Electrons and Peter Grünberg Institute, Forschungszentrum Jülich, 52425 Jülich, Germany

‡Author to whom correspondence should be addressed: kyulee@uni-mainz.de

**Abstract**

The recent emergence of magnetic van der Waals materials allows for the investigation of current induced magnetization manipulation in two dimensional materials. Uniquely, Fe$_3$GeTe$_2$ has a crystalline structure that allows for the presence of *bulk* spin-orbit torques (SOTs), that we quantify in a Fe$_3$GeTe$_2$ flake. From the symmetry of the measured torques, we identify the current induced effective fields using harmonic analysis and find dominant *bulk* SOTs, which arise from the symmetry in the crystal structure. Our results show that Fe$_3$GeTe$_2$ uniquely can exhibit *bulk* SOTs in addition to the conventional interfacial SOTs enabling magnetization manipulation even in thick single layers without the need for complex multilayer engineering.

The discovery of magnetic van der Waals crystals that retain magnetic order in the two-dimensional limit [1–4] opens up the investigation of their magnetic properties and implementation in spintronic devices. To this end, the efficient control of the magnetic state is essential. Spin-orbit torques (SOTs) provide the opportunity of an electrical control of the magnetization [5]. Linking magnetic van der Waals materials with spin-orbit torques exhibits the potential of current induced magnetization manipulation in two dimensions, thus enabling fast and low power spintronic devices. So far reports of current-induced switching of 2D magnets have been based on interfacial SOTs that rely on complex multilayers requiring a small thickness of the magnetic layer thus limiting thermal stability. [6–8] One of the promising magnetic van der Waals crystals is $Fe_3GeTe_2$, which has been shown to exhibit a strong perpendicular magnetocrystalline anisotropy [9], which is still present in an atomic monolayer [10] combined with the Dzyaloshinskii-Moriya interaction stabilizing skyrmions [11–13]. Furthermore, of the van der Waals materials it shows one of the highest bulk Curie temperatures of ~ 225 K [9,14] which can be increased up to room temperature by ionic gating [15].

While switching due to interfacial spin-orbit torques in $Fe_3GeTe_2$ has been studied in multilayer structures, Johansen et al. [16] recently predicted that a possible bulk SOT could be present due to the symmetry of the monolayer crystalline structure. Since $Fe_3GeTe_2$ is a metal, the combination of the perpendicular magnetic anisotropy with this theoretically predicted *bulk* SOTs could potentially enable simple new devices: *bulk* SOTs are efficient for thick samples that have a better magnetic thermal stability than the thin flakes necessary for interfacial SOTs, while also drastically simplifying single layer device engineering. However, so far the symmetry analysis has only indicated that *bulk* SOTs are by symmetry allowed [16], but the torques in single $Fe_3GeTe_2$ layers have not been quantified. Although there have been demonstrations of bulk-type SOTs in 3D magnets such as $L1_0$ FePt and Co-Tb, the origin and the behavior of the SOTs are different from the special *bulk* SOTs that are only present in materials with particular crystalline structure as studied here in $Fe_3GeTe_2$ [17,18]. So there is a clear need to check the presence, identify the origin and quantify the amplitude of the torques as key steps forward.

In this work, we fabricate a $Fe_3GeTe_2$ device with transverse Hall contacts and measure the higher harmonic Hall voltages in order to derive the current induced effective spin-orbit fields. We analyze the symmetries and amplitudes of the torques to understand the bulk and interfacial contributions to the torques and ascertain the temperature dependence highlighting exceptionally large bulk torques in this system.

In a first step, polycrystalline $Fe_3GeTe_2$ was synthesized from the elements Fe, Ge and Te in a solid-state reaction. A stoichiometric mixture was vacuum-sealed in a quartz glass ampoule, heated up to 625 °C with 100 K/h, and kept at this temperature for 60 hours before cooling to room temperature with 100 K/h. [19] Phase-purity of the gray powder was ensured by X-ray diffraction. In a second step, single crystals of an average size of 1.5 x 1.5 x 0.5 mm$^3$ were grown from the polycrystalline powder via a chemical vapor transport with iodine as the transport agent in a temperature gradient from 750 °C to 700 °C for one week. [20] The atomic structure was verified using high-resolution scanning transmission electron microscopy in Fig. S1a (see also Supplementary Note 1). We confirmed the stoichiometric composition of the $Fe_3GeTe_2$ by energy dispersive X-ray spectroscopy within an error of about 5% (see Supplementary Note 2). From the grown bulk crystals, $Fe_3GeTe_2$ flakes were manually exfoliated and placed on an undoped naturally oxidized silicon substrate. The thickness of the flake was measured with an atomic force microscope. By using electron beam lithography, gold contacts were fabricated. The final device can be seen in the inset of Fig. 1a. The flake used in the following study is measured to be 35 nm thick. Assuming the current flows in between the transverse contacts, the cross-section area of the investigated flake is estimated to A = 1.75 × 10$^{-13}$ m$^2$, which is the value used to calculate the current densities.

In order to measure the SOTs, higher harmonic Hall measurements [21] are performed. The measurements were done with the configuration shown in Fig. 1a where the current was applied in the $x$ direction and the Hall voltage $U_H$ was measured in the $y$ direction. The perpendicular magnetic anisotropy of the flake was confirmed by the anomalous Hall measurements for different temperatures as shown in Fig. 1b. While at low temperature abrupt switching is seen, a multi-step switching appears above 175 K near $T_c$ which shows a formation of domains and even skyrmions as observed by TEM in Fig. S4 (see supplementary Note 4). At various temperatures and applied alternating currents $j_C$, the first ($U_H^{1st}$) and second harmonic ($U_H^{2nd}$) Hall voltages are recorded using lock-in amplifiers. For each combination of temperature and current, a magnetic field **B** is applied in the plane to tilt the magnetization **M**. Thereby, the external field is either aligned with the longitudinal current direction ($\Phi_B = 0°$) or perpendicular to it ($\Phi_B = 90°$). A small z-component prevents multi-domain nucleation, so that the polar angle of the external field ranges between $80° \leq \theta_B \leq 83°$. To take care of heating effects, the second harmonic signal is corrected according to the method outlined in [21].

By consideration of the anomalous and planar Hall effect, the Hall voltage in general can be expressed by [21]:

$$U_H = R_{AHE}\, j_C\, \cos(\theta) + R_{PHE}\, j_C\, \sin^2(\theta)\, \sin(2\Phi). \qquad (1)$$

$R_{AHE}$ and $R_{PHE}$ are the anomalous and planar Hall effect coefficients and $j_C$ is the applied current. Expanding equation (1) to first order of the applied current with $j_C = j_0 \cos(\omega t)$ results in:

$$U_H = U_H^{0th} + U_H^{1st} \cos(\omega t) + U_H^{2nd} \cos(2\omega t), \qquad (2)$$

$$U_H^{1st} = R_{AHE}\, j_0\, \cos(\theta_0) + R_{PHE}\, j_0\, \sin^2(\theta_0)\, \sin(2\Phi_0), \qquad (3)$$

$$U_H^{0th} = U_H^{2nd} = j_0^2/2\, (R_{AHE} - 2\, R_{PHE}\, \cos(\theta_0)\, \sin(2\Phi_0))\, \left.\frac{\partial \cos(\theta)}{\partial B}\right|_{\theta_0}\, \frac{1}{\sin(\theta_B - \theta_0)}\, b^{\theta}_{SOT} +$$

$$j_0^2/2\, R_{PHE}\, \sin^2(\theta_0)\, \frac{2\cos(2\Phi_0)}{B \sin(\theta_B)}\, b^{\Phi}_{SOT}. \qquad (4)$$

Here, $b^{\theta}_{SOT}$ and $b^{\Phi}_{SOT}$ are the derivatives of the $\theta$- and $\Phi$-components of the effective spin-orbit fields with respect to the current density and $\theta_0$ and $\Phi_0$ are the equilibrium angles of the material's magnetization. In the following, we assume that the azimuthal angles of the magnetization and the magnetic field are equal: $\Phi \equiv \Phi_0 = \Phi_B$. Due to the measurement configuration with $\Phi = 0°/90°$, all terms including $\sin(2\Phi_0)$ become zero in equation (3) and (4). In [16] a description of the current induced effective spin-orbit field has been derived in particular for the Fe$_3$GeTe$_2$ crystal structure:

$$\mathbf{B_{SOT}} = \Gamma_0 \, J_C \, (m_x \, \mathbf{e_x} - m_y \, \mathbf{e_y}). \qquad (5)$$

$\Gamma_0$ is a parameter, which represents the strength of the spin-orbit torques, $J_C$ is the applied current density, $m_i$ is the i$^{th}$ component of the magnetization unit vector and $\mathbf{e_x}$ and $\mathbf{e_y}$ are the x- and y-unit vectors. We see that these torques lead to a canting of the spins into the plane of the sample and can facilitate switching by reducing the switching energy barrier. The canting lends itself naturally to detection by higher harmonic Hall measurements. By transforming equation (5) to spherical coordinates and considering the measurement configuration $\Phi = 0°/90°$, we find that the spin-orbit field only comprises a $\theta$-component:

$$B^\theta{}_{SOT} = \Gamma_0 \, J_0/2 \, \sin(2\theta_0) \, \cos(2\Phi). \qquad (6)$$

$J_0 = j_0/A$ is the current density. For this reason, $b^\Phi{}_{SOT}$ in equation (4) is also zero and the first and second harmonic Hall resistances can finally be formulated to:

$$R_H{}^{1st} = R_{AHE} \cos(\theta_0), \qquad (7)$$

$$R_H{}^{2nd} = j_0/2 \, R_{AHE} \left.\frac{\partial \cos(\theta)}{\partial B}\right|_{\theta_0} \frac{1}{\sin(\theta_B - \theta_0)} b^\theta{}_{SOT}. \qquad (8)$$

Consequently, by measuring $R_H{}^{2nd}$, $\theta_0$ and $R_{AHE} \left.\frac{\partial \cos(\theta)}{\partial B}\right|_{\theta_0} = \left.\frac{\partial R_H{}^{1st}}{\partial B}\right|_{\theta_0}$, we derive the current induced effective SOTs $b^\theta{}_{SOT}$. In Fig. 2 the first and second harmonic Hall resistances after correction for heating effects are plotted as a function of the applied magnetic field when the magnetic field applied in the $\Phi = 0°$ and 90° direction. Note that the field dependence of the second harmonic signal in the $\Phi = 0°$ and 90° configurations is found to be odd and even, respectively.

The next step is to check the nature of the torques by measuring their symmetry. To obtain the polar angular dependence ($\theta_0$), we plot in in Fig. 3a the current-induced effective spin-orbit torques $b^\theta_{SOT}$ as a function of the applied magnetic field, which corresponds to a certain polar angle. Data points corresponding to smaller external fields are omitted, due to the fact that the term $\frac{\partial R_H^{1st}}{\partial B}\big|_{\theta_0}$ diverges near the switching region. In Fig. 3b the same $b^\theta_{SOT}$ data is plotted as a function of the extracted polar magnetization angle $\theta_0$. To demonstrate the odd-symmetry dependence of the damping-like effective field geometry ($\Phi = 0°$) and check that they overlap, we invert the data points corresponding to negative applied fields. The field-like effective field geometry ($\Phi = 90°$) values range between -2 and 4 mT/$10^{11}$ Am$^{-2}$ with the highest values at smaller $\theta_0$ angles. The damping-like effective fields decrease in magnitude from -8 to -3 mT/$10^{11}$ Am$^{-2}$ with increasing angle. Above $\theta_0 > 45°$ the absolute value of $b^\theta_{SOT}$ increases again up to -4 mT/$10^{11}$ Am$^{-2}$.

The key step now is to clarify the origin of these torques to check if they are of bulk origin as predicted. A first observed key feature that allows us to identify the bulk origin is the opposite behaviour of the effective spin-orbit fields for $\Phi = 0°$ and $\Phi = 90°$ as predicted by equation (6) due to the cos($2\Phi$) term that yields +1 for $\Phi = 0°$ and –1 for $\Phi = 90°$.

Secondly from the $\theta_0$ dependence we can identify a dominating bulk origin. We can fit equation (6) for pure *bulk* SOTs to the data in Fig. 3b. To check if additionally interfacial torques play a role, the fit equation was extended to take into account an additional interfacial SOTs [21]. Accordingly, we fit our data with a combination of bulk and interfacial SOTs [16,21]:

$$b^\theta_{SOT} (\Phi = 0°) = \Gamma_0^\parallel/2 \sin(2\theta_0) + T_0^\parallel + T_2^\parallel \sin^2(\theta_0) \tag{9}$$

$$b^\theta_{SOT} (\Phi = 90°) = -\Gamma_0^\perp/2 \sin(2\theta_0) - \cos(\theta_0) (T_0^\perp + T_2^\perp \sin^2(\theta_0)) \tag{10}$$

Where $T_i^{\parallel}$ and $T_i^{\perp}$ are the $i^{th}$ order components of the longitudinal and transverse components of spin-orbit torques from the interface. Fig. 3b shows the resulting fit of the interfacial spin-orbit torques.

To finally check if the key feature of *bulk* SOTs, namely the opposite behavior for $\Phi = 0°$ and $\Phi = 90°$ is universally present or a random occurrence by chance at 175 K (the data shown in Fig. 3), we investigate the temperature dependence. By extracting $\Gamma_0$ from each fit, the fundamental magnitude of *bulk* SOTs in Fe$_3$GeTe$_2$ can be determined as a function of temperature as shown in Fig. 4. As visible in the temperature dependence, the behavior of $\Gamma_0 *\cos(2\Phi)$ exhibits consistently at all temperatures opposite behavior for $\Phi = 0°$ and $\Phi = 90°$ in line with the prediction for *bulk* SOTs. Note that the fits for $\Gamma_0$ were done independently for $\Phi = 0°$ and $\Phi = 90°$ so that one could easily determine if the temperature dependence were qualitatively different for both orientations. However here we see that for both orientations the largest values for $\Gamma_0$ occur at the lowest temperatures and then decrease to lower values for temperatures up to 150K before increasing again and peaking around 215 K. Note that the reduction of $\Gamma_0$ at even higher temperatures approaching the Curie temperature (230 K) is expected as the magnetic order is lost and at elevated temperature close to the Curie temperature inhomogeneous properties and domain formation can make the analysis less robust. The complex behavior of $\Gamma_0$ at elevated temperatures could thus be related to the multi-step switching indicating domain formation shown in Fig. 1b. We note that at higher temperatures the error bars are smaller as the magnetization can be tilted with our maximum available vector field of 5T to higher angles and a wider $\theta_0$ angle range can be investigated and fitted since the anisotropy decreases with temperature.

In particular, the data shows that at low temperatures we measure very high values of $b^{\theta}_{SOT}$ of more than 50 mT/10$^{11}$ Am$^{-2}$. Together with the very high interfacial SOTs found in Fe$_3$GeTe$_2$/Pt structures [7,8], this bodes well for efficient switching of the magnetization in this material by combined bulk and interfacial torques.

In the following, we discuss the possible origins of the SOTs that we measure. In [16] a SOT mechanism related to a broken inversion symmetry of the structure is introduced. While the polar angular dependences found at the different temperatures indicate a dominating bulk origin of the torques, we see that the temperature dependence does show some deviations from the strict proportionality expected from eq. 6 and the absolute values for $\Phi = 0°$ and $\Phi = 90°$ shown in Fig. 4 do not always fully coincide. This indicates that additional higher order torques that can for instance be induced by uniaxial strain can play a role, highlighting the breadth of new torques that can contribute due to our identified bulk mechanisms. Furthermore, we see that fits of the polar angular dependence (Fig. 3) indicate additional interfacial torques beyond the intrinsic bulk torque can be present. To check why interfacial torques can occur, even though the $Fe_3GeTe_2$ device that is measured is in principle a bulk device with a thickness of 35 nm, where no net interfacial torques due to the spin Hall effect or the inverse spin-galvanic effect [5,22] are expected, we consider surface oxidation. Overall, this device has been exposed to air < 12 hours and literature reports a natural oxidation layer on an exfoliated flake within a time scale of 14 hours [23,24]. The presence of an oxide layer on the surface exposed to air is confirmed by TEM (see supplementary material fig. S3) and thus interfacial SOTs can appear. So we find that in our $Fe_3GeTe_2$ device clear evidence for a theoretically reported *bulk* SOT based on the crystal symmetry breaking [16] and on the other hand an additional interfacial SOT are enabled by local surface oxidation. A third possible contribution to the measured spin-orbit torques could be Oersted fields, which are additional magnetic fields, which arise due to the current flow and can mimic a field-like torque symmetry. Assuming that the $Fe_3GeTe_2$ flake is a homogeneous conductor, the Oersted field is zero in the center of the flake and rises in magnitude at the edges, pointing counterclockwise in the *yz*-plane [25]. Thus, it points in opposite *y*-directions at the top and bottom of the flake and cancels to zero. If we assume that the $Fe_3GeTe_2$ flake has become a heterogeneous conductor due to possible interfaces, the current flow in the *z*-direction becomes asymmetric and hence also the Oersted field. In the less conducting areas, this Oersted field can be estimated by $H_{Oe} = \mu_0 j_C/(2\pi r)$ according to the Biot-Savart law for an infinite long straight conductor with r the distance from the conductor. Therefore, exactly at the interfaces to less conducting areas e.g. at the top of the flake the Oersted field becomes maximal. However, given that we probe the bulk of the flake, the contribution of the Oersted field will be negligible compared to the measured torque values.

In order to quantify the *bulk* SOT from a theoretical perspective, we employ the microscopic first-principles framework to compute the anti-damping SOT within the Kubo linear response theory for the $Fe_3GeTe_2$ bulk crystal (details are given in the Supplementary Note 5). As bulk $Fe_3GeTe_2$ maintains inversion symmetry the SOT vanishes globally [26]. However, each layer separately may exhibit a non-vanishing SOT which may lead to the non-vanishing effect observed experimentally. If we decompose the unit cell into the A and B layers of $Fe_3GeTe_2$ (Fig. S4 of the supporting material) the top and bottom Fe atoms of the A-layer experience an equal in magnitude but opposite in sign SOT for an out-of-plane magnetization. However, once the magnetization is tilted away from the out-of-plane direction the SOT for each layer does not cancel out separately. From first principles, the estimated magnitude of the SOT per layer with a magnetization angle of $\theta = 30°$ from the z axis, and $\varphi = 55°$ from the current direction of is 2.19 $mT/(10^{11} Am^{-2})$ and of similar order to that found experimentally (Fig. 3b at $\theta_0$=30°, $\Phi = 55°$).

In summary, we have measured SOTs in a pure $Fe_3GeTe_2$ flake with very large magnitudes of more than 50 mT/$10^{11}$ Am$^{-2}$. From a symmetry analysis we can identify the predicted *bulk* SOTs that result from the particular crystalline structure of the $Fe_3GeTe_2$ that we determine by TEM imaging. In addition, we find that interfacial SOTs are present that result likely from surface effects such as observed oxidation. *Ab initio* calculations confirm that the layer resolved *bulk* SOT is of the same order of magnitude as the experiment. Our findings that the *bulk* SOTs that are a unique property of certain van der Waals materials such as $Fe_3GeTe_2$ yield very efficient magnetization manipulation due to high effective fields combined with the possibility of simple device design with just a single material without any additional materials and layers lays the foundations for a new paradigm of 2D materials spin-orbitronic devices.

**Acknowledgements**

The work at JGU Mainz was funded by the Deutsche Forschungsgemeinschaft (DFG, German Research Foundation) – TRR 173 – 268565370 (projects A01 and B02), the EU (FET-Open grant agreement no. 863155 (s-Nebula), ERC Synergy grant agreement no. 856538 (3D MAGiC), and the Research Council of Norway (QuSpin Center 262633). T.S., D.G., and Y.M. gratefully acknowledge the Jülich Supercomputing Centre for providing computational resources and Deutsche Forschungsgemeinschaft (DFG, German Research Foundation) – TRR 173 – 268565370 (project A11), TRR 288 – 422213477 (project B06).

Fig. 1. (a) Scheme of the 2nd harmonic Hall measurement. An alternating current is injected along the *x*-direction, while the transverse 1st and 2nd harmonic Hall voltage $U_{trans}$ is measured via a lock-in amplifier. In the inset an optical microscope image of the final device is depicted. (b) The hysteresis loops of $Fe_3GeTe_2$ at different temperatures with the magnetic field applied in the z direction.

Fig. 2. Examples of the 1st and 2nd harmonic Hall resistances as a function of the applied magnetic field along the *x*-direction $\Phi = 0°$ (a) and y-direction $\Phi = 90°$ (b) at a temperature of 100 K with a polar magnetic field angle of $\theta_B = 82°$. The applied current density is $4.1 \times 10^{10}$ $Am^{-2}$.

Fig. 3. The derivative of the $\theta$ component of the current induced effective field is shown as a function of the externally applied magnetic field (a) and the polar magnetization angle $\theta_0$ (b) at a temperature of 175 K with a polar magnetic field angle of $\theta_B = 82°$. The applied current density is $3.7 \times 10^{10}$ $Am^{-2}$. In (b) the data for $\Phi = 0°$ and negative applied fields has been inverted. The solid lines are fits according to equation (9) and (10).

Fig. 4. The extracted *bulk* spin-orbit torque parameter $\Gamma_0 * \cos(2\Phi)$ as a function of the temperature showing the opposite sign for the $\Phi = 0°$ and $\Phi = 90°$ data over the full temperature range.

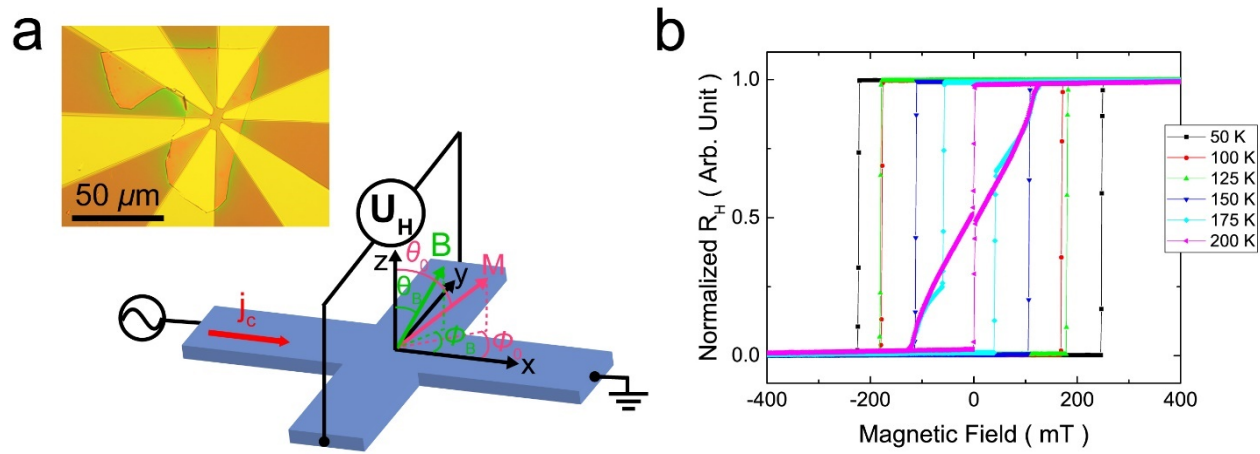

Figure 1

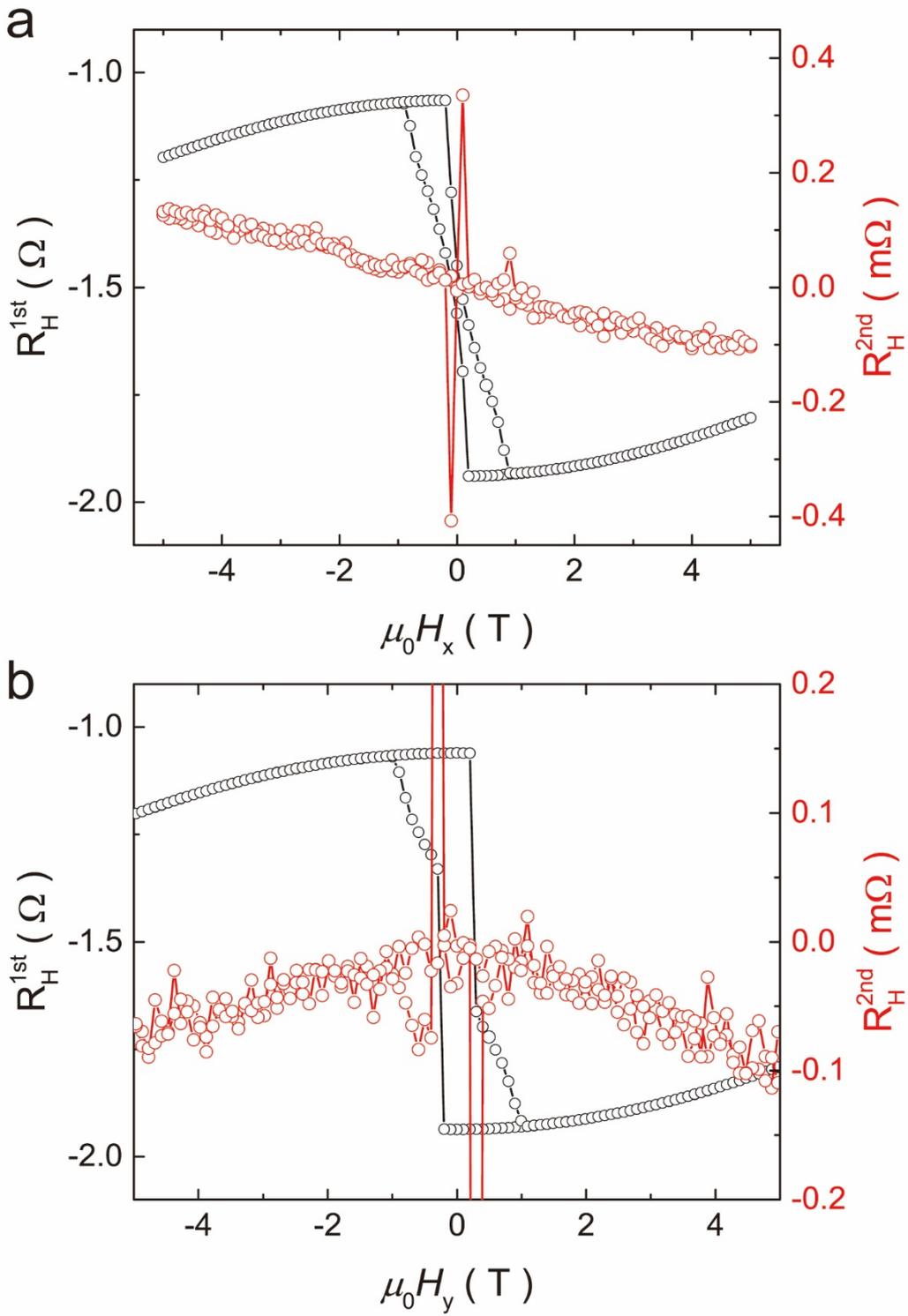

Figure 2

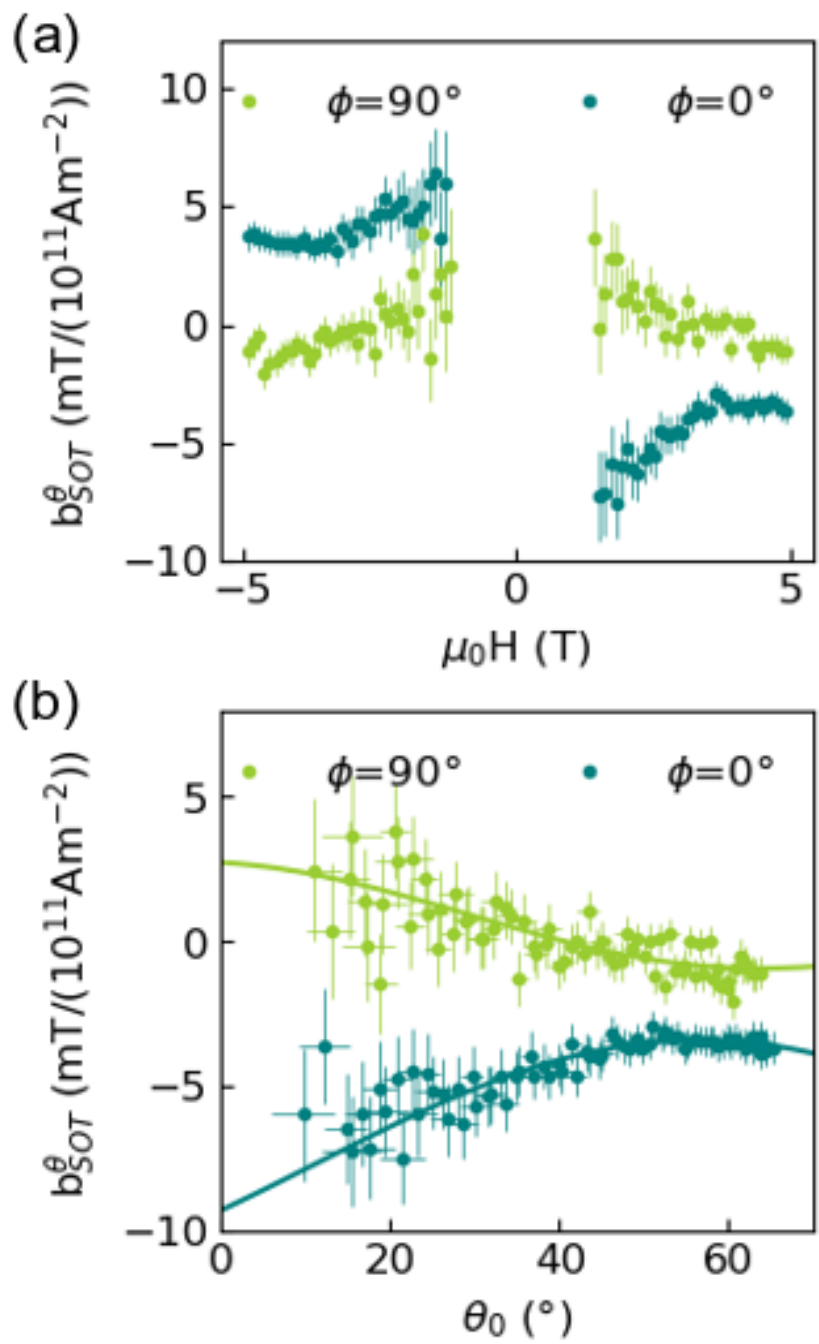

Figure 3

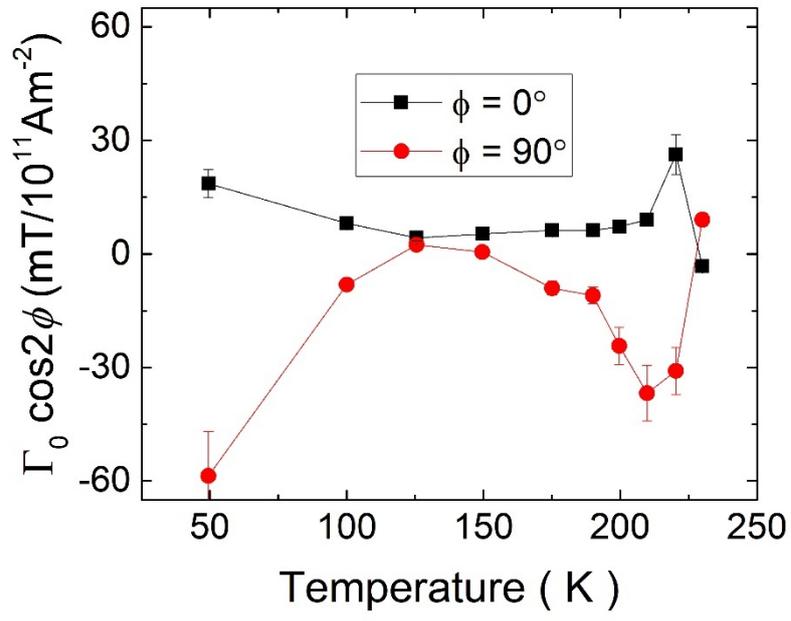

Figure 4